\documentclass[12pt,english,oneside]{amsart}
\usepackage[T1]{fontenc}
\usepackage[latin1]{inputenc}
\usepackage{amsthm}
\usepackage{amssymb}
\usepackage{graphicx}

\makeatletter
\numberwithin{equation}{section}
\numberwithin{figure}{section}
\theoremstyle{plain}
\newtheorem{thm}{\protect\theoremname}
  \theoremstyle{definition}
  \newtheorem{defn}[thm]{\protect\definitionname}

\usepackage{color}
\usepackage{epsfig}

\makeatletter

\usepackage{amsfonts}

\makeatother

\makeatother

\usepackage{babel}
  \providecommand{\definitionname}{\inputencoding{latin9}Definition}
\providecommand{\theoremname}{\inputencoding{latin9}Theorem}

\begin{document}

\title{Scalar field on non-integer dimensional spaces }

\author{R. Trinchero}

\date{20 December 2011}
\begin{abstract}
Deformations of the canonical spectral triples over the $n-$dimensional
torus are considered. These deformations have a discrete dimension
spectrum consisting of non-integer values less than $n$. The differential
algebra corresponding to these spectral triples is studied. No junk
forms appear for non-vanishing deformation parameter. The action of
a scalar field in these spaces is considered, leading to non-trivial
extra structure compared to the integer dimensional cases, which does
not involve a loss of covariance. One-loop contributions are computed
leading to finite results for non-vanishing deformation.
\end{abstract}

\address{Instituto Balseiro y Centro Atómico Bariloche.}

\email{trincher@cab.cnea.gov.ar}

\keywords{Dimensional regularization, non-commutative geometry, non-integer
dimensions. }

\thanks{R.T. is supported by CONICET.}

\maketitle

\section{Introduction}

The dimension of a space is a basic concept of particular relevance
both in nature and in mathematics. Non-commutative geometry\cite{con}\cite{madore,landi,gravarillfigue}
provides a generalization of classical geometry. In particular, it
includes a definition of dimension that allows for complex non-integer
values\cite{cm}. A motivation for this definition and a series of
very interesting examples of geometries with non-integer dimensions
has been given in relation to the study of fractal sets in this geometrical
setting(\cite{con},\cite{lap06} and references therein). 

The motivation for this work comes from a different subject. In the
realm of quantum field theory(QFT), the widely employed dimensional
regularization technique\cite{coll} provides a hint that non-integer
dimensional spaces could be of relevance there. This technique is
employed in QFT as a means to regularize divergent integrals appearing
in perturbation theory, being preferred in the regularization of gauge
theories since it preserves gauge invariance. The technique essentially
consists in considering the analytical continuation in the number
of dimensions for the surface of a d-dimensional sphere, a quantity
that appears in the calculation of the above-mentioned integrals.
The general question to be addressed in this work is whether a suitable
well-defined differential geometry can be found that makes sense for
non-integer dimensions and reduces to the canonical one for the integer
case%
\footnote{A preliminary study of this question in the 1-dimensional case appears
in \cite{rt}%
}. In the affirmative case the natural question to ask is, what does
a field theory defined in such a space look like?. More precisely,
the idea is to take a field theory defined purely in geometrical terms
and repeat the construction in the deformed case. The output of that
procedure is by no means obvious since, as will be seen in subsequent
sections, the differential algebra is qualitatively different between
the integer and non-integer case, and such a change reflects directly
in the action of the field theory. For the case of the field theory
of a scalar field considered in section 6, the resulting theory is
of a novel type. This theory, in spite of reflecting its non-commutative
origin, does not involve a breakdown of covariance, as happens in
the so-called non-commutative field theories\cite{dougnekra}.

The salient features and results of this work are summarized as follows,
\begin{itemize}
\item Spectral triples are considered that differ from the canonical ones
only in the choice of the Dirac operator.
\item The dimension spectrum of these triples consists of a discrete set
of real values less than the dimension of the canonical triple.
\item The differential of a zero form is not a multiplicative operator.
\item There are no junk forms for a non-zero deformation parameter.
\item The action of a scalar field contains derivatives of any order and
involves an integration over the co-sphere.
\item In spite of the \textquotedbl{}non-commutativeness\textquotedbl{}
of the differential algebra, there is no loss of covariance involved
in the field theory mentioned above.
\item The calculation of the tadpole diagram and a loop involving two free
propagators, show that for non-zero deformation these diagrams give
a finite result, showing neither ultraviolet nor infrared singularities.
The last singularities being ruled out by the appearance of a mass
term whose coefficient vanishes when the deformation parameter goes
to zero.
\item Ultraviolet power counting and comparison with dimensional regularization,
indicate that the perturbation theory obtained from the action mentioned
above leads to finite contributions for a non-zero deformation parameter.
\end{itemize}
This paper is organized as follows. Section 2 describes the spectral
triple to be considered. In section 3 the corresponding dimension
spectrum is computed. The differential of a 0-form is considered in
section 4. Section 5 considers the calculation of the action for a
complex scalar field. Section 6 presents the one-loop computations
and Section 7 contains conclusions and the schematic description of
further research motivated by the present work. In addition, two appendices
are included, Appendix A showing the absence of junk forms, and Appendix
B, which contains the calculation of the Wodzicki residue involved
in the definition of the above-mentioned action.

\section{The Dirac operator}

The differential algebra derived from the canonical spectral triple
involving functions over a manifold $M$ reduces to the usual exterior
differential algebra over $M$. The spectral triples to be considered
in this work differ from the canonical ones only in the choice of
the Dirac operator. More precisely, the triples $(\mathcal{A},\mathcal{H},D_{\alpha})$
are considered, where,
\begin{itemize}
\item $\mathcal{A}$ is the commutative $C^{*}$-algebra of smooth functions
over the $n\mbox{-dimensional}$ torus $T^{n}\;\; n\in\mathbb{N}$.
\item $\mathcal{H}$ is the Hilbert space of square integrable sections
of a spinor bundle over $T^{n}$.
\item $D_{\alpha}:\mathcal{H}\to\mathcal{H}$ is a self-adjoint linear operator
to be defined below. 
\end{itemize}
The usual Dirac operator over a $n$ dimensional torus $T^{n}$ is
given by,
\[
D=i\gamma\cdot\partial=i\gamma_{\mu}\partial_{\mu}\;\;,\gamma_{\mu}=\gamma_{\mu}^{\dagger}\,,\;\;\gamma_{\mu}\gamma_{\nu}+\gamma_{\nu}\gamma_{\mu}=2\delta_{\mu\nu}\;,\mu,\nu=1,\cdots,n
\]
this operator is not positive definite. Indeed since,
\[
D^{2}=-\Delta=-\partial_{\mu}\partial_{\mu}
\]
denoting by $\lambda\geq0$ an eigenvalue of $D^{2}$, then $\pm\sqrt{\lambda}$
will be eigenvalues of $D$. 

In this work the usual Dirac operator will be replaced by $D_{\alpha}$
given below. One of the motivations for this choice is to obtain a
dimension spectrum with non-integer real values. This could be done
in many ways, for example, choosing,
\[
D_{a}=D|D^{2}|^{-\frac{(1-a)}{2}}\;\;,a\in\mathbb{R},\;1>a>0
\]
this operator leads to a dimension spectrum%
\footnote{See the next section for the definition of dimension spectrum\cite{cm}.%
} which consists in a single value given by $z=\frac{n}{a}$. However,
it is not well-behaved in the infrared. In order to improve its infrared
properties and have the same behavior in the ultraviolet, the following
operator will be considered in this work,
\[
D_{\alpha}=D(1+D^{2})^{-\alpha},\;\alpha>0
\]
the power appearing in this last equation being defined by,

\begin{equation}
(1+D^{2})^{-\alpha}=\frac{1}{\Gamma(\alpha)}\int_{0}^{\infty}d\tau\,\tau^{\alpha-1}e^{-\tau(1+D^{2})}\label{eq:power}
\end{equation}
 Thus the Dirac operator to be considered is,
\[
D_{\alpha}=\frac{1}{\Gamma(\alpha)}\int_{0}^{\infty}d\tau\,\tau^{\alpha-1}D(\tau)\;\;,D(\tau)=e^{-\tau(1+D^{2})}D
\]
this operator is self-adjoint in $\mathcal{H}$, with compact resolvent,
and such that the differential of any $a\in\mathcal{A}$ is bounded.
This last condition is ensured by the choice $\alpha\geq0$, as can
be readily shown using the expression for the differential of section
4. Therefore, the triple fulfills all the properties required for
it to be a spectral triple.

\section{Dimension spectrum}

The definition of dimension spectrum of a spectral triple is briefly
reviewed. 
\begin{defn}
{[}Connes-Moscovici{]}\label{cm} \emph{Discrete dimension spectrum.}
A spectral triple $(\mathcal{A},\mathcal{H},D)$ has discrete dimension
spectrum $Sd$ if $Sd\subset\mathbb{C}$ is discrete and for any element
$b$ in the algebra%
\footnote{The definition of the algebra $\mathcal{B}$ is the following. Let
$\delta$ denote the derivation $\delta:L({\mathcal{H}})\to L({\mathcal{H}})$
defined by, 
\begin{equation}
\delta(T)=[|D|,T]\qquad,T\in L({\mathcal{H}})\label{15}
\end{equation}

The algebra $\mathcal{B}$ is generated by the elements,
\begin{equation}
\delta^{n}(\pi(a)),\; a\subset\mathcal{A},\;\; n\geq0\;(\delta^{0}(\pi(a))=\pi(a))\label{16}
\end{equation}
} $\mathbf{\mathcal{B}}$ the function, 
\begin{equation}
\zeta_{b}^{D}(z)=Tr[\pi(b)\:|D|^{-z}]\label{17}
\end{equation}
 extends holomorphically to $\mathbb{C}/Sd$. 

The interpretation of these poles is that each of them gives the dimension
of a certain piece of the whole space.
\end{defn}
In order to apply this definition to the spectral triples considered
in this work, it is useful to note that,
\begin{eqnarray}
|D_{\alpha}|^{-z} & = & |D|^{-z}(1+|D|^{2})^{\alpha z}=|D|^{-z}\sum_{k=0}^{\infty}\left(\begin{array}{c}
\alpha z\\
k
\end{array}\right)|D|^{2(\alpha z-k)}\nonumber \\
 & = & \sum_{k=0}^{\infty}\left(\begin{array}{c}
\alpha z\\
k
\end{array}\right)|D|^{2((\alpha-\frac{1}{2})z-k)}\label{eq:newbinob-1}
\end{eqnarray}
where  Newton's binomial formula has been employed. From the definitions
above it is clear that,
\begin{equation}
\zeta_{b}^{D_{\alpha}}(z)=\sum_{k=0}^{\infty}\left(\begin{array}{c}
\alpha z\\
k
\end{array}\right)\zeta_{b}^{D}(2(k-(\alpha-\frac{1}{2})z))\label{eq:zeta}
\end{equation}
where the binomial coefficients are given by,
\[
\left(\begin{array}{c}
\alpha z\\
k
\end{array}\right)=\frac{\alpha z(\alpha z-1)\cdots(\alpha z-k+1)}{k!}\;\;,\left(\begin{array}{c}
\alpha z\\
0
\end{array}\right)=1
\]
The zeta functions appearing in the r.h.s. of (\ref{eq:zeta}) are
the ones corresponding to the canonical spectral triple. Thus, since
for the canonical spectral triples the corresponding zeta functions
have a single simple pole at its argument equal to $n$, then $\zeta_{b}^{D_{\alpha}}(z)$
has  simple poles at,
\[
z=\frac{n-2k}{1-2\alpha}\;\;,k=0,1,2,\cdots
\]
these values of $z$ are therefore the dimension spectrum of the spectral
triple considered in this work.

\section{The differential}

The differential of a 0-form $f$ is given by,
\begin{eqnarray}
df & = & [D_{\alpha},f]=\frac{1}{\Gamma(\alpha)}\int_{0}^{\infty}d\tau\,\tau^{\alpha-1}df(\tau)\label{eq:dif}\\
df(\tau) & = & [D(\tau),f]\;\;.D(\tau)=U(\tau)D\;\;,U(\tau)=e^{-\tau(1+D^{2})}
\end{eqnarray}
thus when applied to an element $\phi$ of $\mathcal{H}$, $df(\tau)$
is given by,
\begin{eqnarray}
df(\tau)\,\phi & = & [D(\tau),f]\:\phi=U(\tau)[(Df)\,\phi+fD\phi]-f\, U(\tau)D\phi\nonumber \\
 & = & \left[U(\tau)(Df)+[U(\tau)f-f\, U(\tau)]D\right]\phi\nonumber \\
 & = & U(\tau)\left[(Df)+[f-U(-\tau)f\, U(\tau)]D\right]\phi\label{eq:diftau}
\end{eqnarray}
the second term in the parenthesis of the r.h.s. can be expressed
as,
\begin{equation}
e^{\tau(1+D^{2})}f(x)\, e^{-\tau(1+D^{2})}=f(x-2\tau\partial)\label{eq:mecq}
\end{equation}
this can be easily derived using an analogy with quantum mechanics.
This is done noting that $e^{-\tau(1+D^{2})}$ is, up to a constant,
the imaginary time evolution operator for a free particle of mass
$m=1/2\:$. Thus,
\[
df(\tau)=U(\tau)\left[(Df)-[f(x)-f(x-2\tau\partial)]D\right]
\]
integrating the second line in (\ref{eq:diftau}) as in (\ref{eq:dif})
leads to,
\[
df=(1+D^{2})^{-\alpha}(Df)+[(1+D^{2})^{-\alpha}f-f(1+D^{2})^{-\alpha}]i\gamma\cdot\partial
\]
which clearly shows that when $\alpha\to0$, $df\to i\gamma\cdot\partial f$,
which is the corresponding expression in the canonical case. It is
worth remarking that, as the last equations indicate, this differential
is a non-multiplicative operator for any value of $\alpha\neq0$.
As Appendix A shows, this fact plays an important role in showing
the absence of junk forms.

\section{The scalar field\label{sec:The-scalar-field}}

In this section the part of this space corresponding to the highest
pole will be considered, i.e. for $d=\frac{n}{1-2\alpha}$. The action
for a free scalar field propagating in this space is taken to be,
\[
S=\frac{1}{2}<d\phi,d\phi>
\]
where $\phi$ is a 0-form and the norm in the space forms is given
by%
\footnote{See for example ref.\cite{landi}%
},
\begin{eqnarray}
<\omega,\omega> & = & tr_{\omega}[\omega\omega^{\dagger}|D_{\alpha}|^{-d}]\label{eq:scalfor}
\end{eqnarray}
thus,
\[
S=-tr_{\omega}[d\phi d\phi{}^{*}|D_{\alpha}|^{-d}]
\]
where it was used that $d\phi^{\dagger}=-d\phi^{*}$ and $tr_{\omega}$
denotes the Diximier trace. In the evaluation of this trace it is
important to note that replacing $d=\frac{n}{1-2\alpha}$ in (\ref{eq:newbinob-1})
leads to,
\begin{equation}
|D_{\alpha}|^{-d}=|D_{\alpha}|^{-\frac{n}{1-2\alpha}}=\sum_{k=0}^{\infty}\left(\begin{array}{c}
\frac{\alpha n}{1-2\alpha}\\
k
\end{array}\right)|D|^{-n-2k}\label{eq:newbinob}
\end{equation}
 Therefore $S$ is given by,

\begin{eqnarray*}
S & = & \sum_{k=0}^{\infty}\left(\begin{array}{c}
\frac{\alpha n}{1-2\alpha}\\
k
\end{array}\right)S_{k}\\
S_{k} & = & -tr_{\omega}[d\phi(\tau)d\phi(\tau')^{*}|D|^{-n-2k}]
\end{eqnarray*}
 Noting that,
\begin{eqnarray*}
d\phi & = & [U_{\alpha}D,\phi(x)],\;\; U_{\alpha}=(1+D^{2})^{-\alpha}\\
d\phi^{*} & = & [U_{\alpha}D,\phi^{*}(x)]
\end{eqnarray*}
leads to,
\[
S_{k}=tr_{\omega}\left\{ [U_{\alpha}D,\phi][DU_{\alpha},\phi^{*}]|D|^{-n-2k}\right\} 
\]
Thus replacing the expression obtained in Appendix B for $S_{k}$
leads to,
\begin{equation}
S=-\frac{2^{[\frac{n}{2}]}V_{S^{n-1}}}{n(2\pi)^{n}}\int_{T^{n}}\phi(D^{2}+\frac{\alpha n}{1-2\alpha})(1+D{}^{2})^{-2\alpha}\phi^{*}\label{eq:action}
\end{equation}
where $V_{s^{n-1}}=2\pi^{n/2}/\Gamma(n/2)$ is the area of the $n-1$
dimensional sphere. It is worth noting that in spite of starting with
an action involving no mass term, the fact of working on a non-integer
dimensional space generates effectively such a term as shown by (\ref{eq:action}),
with a coefficient that vanishes in the integer case($\alpha=0$).
In that case (\ref{eq:action}) reduces to the usual action of a mass
less complex scalar field, i.e.,
\[
S_{can}=\lim_{\alpha\to0}S=\frac{2^{[\frac{n}{2}]}V_{s^{n-1}}}{(2\pi)^{n}}\int_{T^{n}}\,\left(\frac{1}{2}\partial_{\mu}\phi(x)\partial_{\mu}\phi^{*}(x)\right)
\]

\section{One loop calculations}

As mentioned in the introduction, the dimensional regularization technique
is a widely employed tool used to make sense of divergences in perturbative
quantum field theory. These divergences appear when calculating the
contribution of Feynman diagrams involving closed loops. Having obtained
a field theory in a non-integer dimensional space, it is natural to
perform the same calculations and see whether the analogous diagrams
are divergent or not. This is the purpose of this section. Two simple
diagrams are considered: the tadpole diagram and a loop involving
two free propagators. In both cases the calculations below show that
the corresponding diagrams give a finite result for $\alpha\neq0$.
The comparison of the results for these diagrams with dimensionally
regularized ones, show that the location of poles in the complex $\alpha$
plane coincides with the one obtained in dimensional regularization.
A simple argument showing that this should be so is obtained by considering
the ultraviolet behavior of the integrals involved, as shown in subsection
6.1. In spite of these similarities, the dimensionally regularized
result and the ones in this non-integer dimensional space are different.
Another important point is that the appearance of the mass term in
(\ref{eq:action}) automatically regulates possible infrared divergences. 

In order to compare with results of standard dimensional regularization
calculations, it is convenient to restore physical units in our calculations.
Unlike common use in physics, where coordinates are assumed to have
dimensions of length, up to this point in this work coordinates and
fields have been taken to be dimensionless. Physical units, in the
natural system of units where action and speed are measured in units
of the Planck constant $\hbar$ and the velocity of light $c$, are
restored by,
\[
x_{P}=\frac{x}{M}\;,\;\;\phi_{P}=M^{\frac{n-2}{2}}\phi
\]
where $x_{P}$ and $\phi_{P}$ denote the dimensionfull quantities
and $M$ is a mass scale. This can be derived recalling the basic
requirement that the action should be dimensionless in natural units.
To show how this works it is noted for example that,
\[
\frac{\partial}{\partial x_{P}}=\frac{1}{M}\,\frac{\partial}{\partial x}\Rightarrow(1+D^{2})=\frac{1}{M^{2}}(M^{2}+D_{P}^{2})
\]
In the subsections below it should be understood that the quantities
involved are dimensionfull, although the subindices $P$ will not
be explicitly written.

\subsection{The tadpole }

According to (\ref{eq:action}) the propagator corresponding to that
action is%
\footnote{In the following expressions the discrete summation over the allowable
momenta is replaced by an integral, this is justified in the limit
where all the radius of the $n$-dimensional torus tend to infinity. %
}, in terms of dimensionfull quantities%
\footnote{In this work $p^{2}>0$ indicates the euclidean positive norm squared
of the vector $p$, this is different from the usual notation in field
theory where after Wick rotating the euclidean momentum is considered
with $p_{E}^{2}<0$.%
},
\[
D(x-y)=\int d^{n}p\frac{1}{(p^{2}+m^{2})(M^{2}+p^{2})^{-\alpha}}e^{-ip\cdot(x-y)}
\]
where,
\[
m^{2}=M^{2}\frac{\alpha n}{1-2\alpha}
\]
 The tadpole diagram is,

\begin{center}
\includegraphics[scale=0.5]{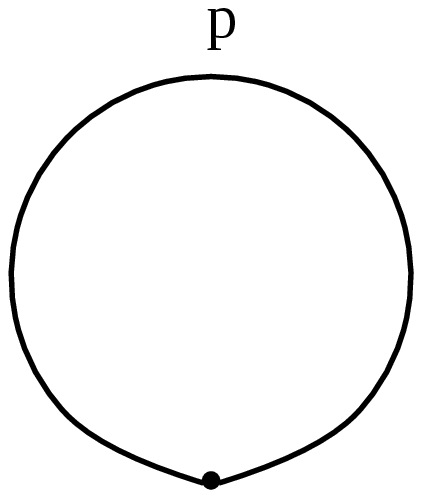}
\par\end{center}

\noindent it corresponds to the following integral,
\begin{equation}
I_{\alpha}^{T}(m)=\int d^{n}p\frac{1}{(p^{2}+m^{2})(M^{2}+p^{2})^{-\alpha}}\label{eq:tadpole}
\end{equation}
It will be shown below that this integral converges for $\alpha\neq0$
and $|\alpha|<1$. A simple argument showing that this should be so
can be given comparing the ultraviolet behavior of $I_{\alpha}^{T}(m)$
and of the corresponding dimensionally regularized integral $I_{d}^{T}(m)$
given by,
\[
I_{d}^{T}(m)=\int d^{d}p\frac{1}{(p^{2}+m^{2})}
\]
the behavior of the integrand in the ultraviolet ($p\to\infty$) is
given by $p^{d-1-2}$, as is well known this dimensionally regularized
integral converges for any $d\neq2,4,6,\cdots$. Next the ultraviolet
behavior of $I_{\alpha}^{T}(m)$ is considered it goes like $p^{n-1-2+2\alpha}$
which coincides with dimensionally regularized case if $d=n+2\alpha$.
The important statement being that considering $\alpha\neq0$ in $I_{\alpha}^{T}(m)$
is equivalent, from the point of view of the ultraviolet behavior
of the integrand, to considering $I_{d}^{T}(m)$ for non-integer $d$.
Of course, the correspondence between ultraviolet behavior of the
integrands does not mean equality of the corresponding integrals,
as is shown by the following calculation.

\noindent In order to evaluate $I_{\alpha}^{T}(m)$, the integral
representation of a power in (\ref{eq:power}) is recalled,
\begin{equation}
A^{-\alpha}=\frac{1}{\Gamma(\alpha)}\intop_{0}^{\infty}d\tau\,\tau^{\alpha-1}e^{-\tau A}\label{eq:potencia}
\end{equation}
this last formula is valid only for negative $\alpha$, which is not
the case of interest here. In what follows, only the case $\alpha<0$
is considered. It will be shown that the final result can be analytically
continued to the case $\alpha>0$. Applying the last formula to (\ref{eq:tadpole})
leads to,
\[
I_{\alpha}^{T}(m)=\int d^{n}p\intop_{0}^{\infty}da\, e^{-a(p^{2}+m^{2})}\frac{1}{\Gamma(-\alpha)}\intop_{0}^{\infty}db\, b^{-\alpha-1}e^{-b(M^{2}+p^{2})}
\]
next, the following change of variables is employed,
\[
z=a+b\;,x=\frac{a}{z}\;\;\Rightarrow a=zx\;,b=z(1-x)
\]
the Jacobian and limits of integration in these new variables implying
that,
\[
\intop_{0}^{\infty}da\, db=-\intop_{0}^{\infty}dz\, z\,\intop_{0}^{1}dx
\]
which leads to,
\[
I_{\alpha}^{T}(m)=\frac{-1}{\Gamma(-\alpha)}\int d^{n}p\intop_{0}^{\infty}dz\, z\intop_{0}^{1}dx\,[z(1-x)]^{-\alpha-1}e^{-z[p^{2}+(1-x)M^{2}+x\, m^{2}]}
\]
in order to make the $p$ integration the following change of variables
is performed,
\[
p\to\tilde{p}=\sqrt{z}p\,\Rightarrow d^{n}p=d^{n}\tilde{p}\, z^{-\frac{n}{2}}
\]
thus,
\[
I_{\alpha}^{T}(m)=\frac{-\pi^{\frac{n}{2}}}{\Gamma(-\alpha)}\intop_{0}^{1}dx\,(1-x)^{-\alpha-1}\intop_{0}^{\infty}dz\, z^{-\alpha-\frac{n}{2}}\, e^{-z[(1-x)M^{2}+x\, m^{2}]}
\]
where use was made of,
\[
\int d^{n}\tilde{p}\, e^{-\tilde{p}^{2}}=\pi^{\frac{n}{2}}
\]
next employing (\ref{eq:power}),
\[
I_{\alpha}^{T}(m)=\frac{-\pi^{\frac{n}{2}}}{\Gamma(-\alpha)}\Gamma(1-\frac{n}{2}-\alpha)\intop_{0}^{1}dx\,(1-x)^{-\alpha-1}[(1-x)M^{2}+x\, m^{2}]^{\alpha+\frac{n}{2}-1}
\]
this last integral can be written in terms of the hypergeometric function
$\,_{2}F_{1}(a,b,c,z)$, i.e.,
\[
I_{\alpha}^{T}(m)=\frac{-\pi^{\frac{n}{2}}\Gamma(1-\frac{n}{2}-\alpha)}{\Gamma(-\alpha)\alpha}(M^{2})^{\alpha+\frac{n}{2}-1}\,_{2}F_{1}(1,1-\alpha-\frac{n}{2},1-\alpha,1-\frac{m^{2}}{M^{2}})
\]
 For illustrative purposes, let us replace $m^{2}=M^{2}\frac{\alpha n}{1-2\alpha}$
by,
\[
\tilde{m}^{2}=m^{2}+m_{0}^{2}
\]
where $m_{0}^{2}$ is a constant additional mass. Taking the limit
$\alpha\to0$ of $I_{\alpha}^{T}(\tilde{m})$ gives,
\[
\lim_{\alpha\to0}\, I_{\alpha}^{T}(\tilde{m})=\pi^{\frac{n}{2}}\Gamma(1-\frac{n}{2})\,(m_{0}^{2})^{\frac{n}{2}-1}
\]
which, upon replacing $n$ by a complex number, is the dimensionally
regularized result of this diagram for a scalar field of mass $m_{0}$.
It is important to note that the analytical properties of both results
are the same, that is so because,
\begin{itemize}
\item $\Gamma(-\alpha)\alpha$ is an analytic function of $\alpha$ for
$|\alpha|<1$.
\item The hypergeometric function  $\,_{2}F_{1}(1,1-\alpha-\frac{n}{2},1-\alpha,1-\frac{m^{2}}{M^{2}})$
is an analytic function of $\alpha$ whenever its third argument is
not equal to $0,-1,-2,\cdots$, i.e. for $\alpha$ not a positive
integer.
\end{itemize}

\subsection{A loop involving two free propagators}

The contribution of the closed loop in the following Feynman diagram,

\begin{center}
\includegraphics[scale=0.5]{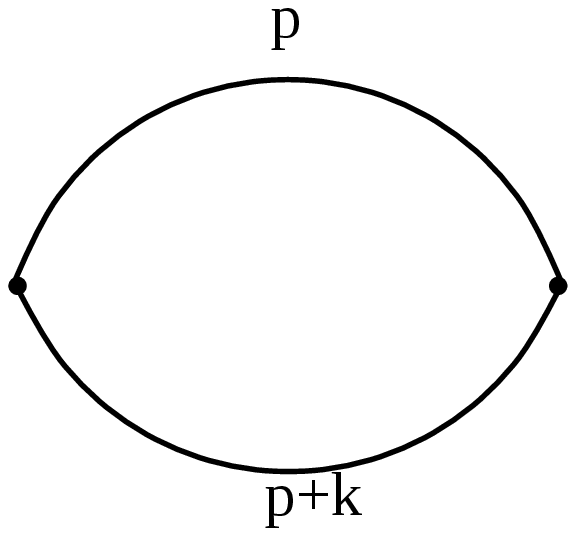}
\par\end{center}

\noindent is given by,
\begin{equation}
I_{\alpha}^{L}(k,m)=\int d^{n}p\frac{1}{(p^{2}+m^{2})(M^{2}+p^{2})^{-\alpha}((p+k)^{2}+m^{2})(M^{2}+(p+k)^{2})^{-\alpha}}\label{eq:ialfa}
\end{equation}
using eq. (\ref{eq:potencia}) the integral $I_{\alpha}^{L}(k,m)$
can be written as follows,
\begin{eqnarray*}
I_{\alpha}^{L}(k,m) & = & \int d^{n}p\frac{1}{\Gamma(-\alpha)^{2}}\intop_{0}^{\infty}da\, db\, d\tilde{a}\, d\tilde{b}\,\tilde{a}^{-\alpha-1}\tilde{b}^{-\alpha-1}\times\\
 &  & e^{E(p,k,a,b,\tilde{a,}\tilde{b})}
\end{eqnarray*}
where,
\begin{eqnarray*}
E(p,k,a,b,\tilde{a,}\tilde{b}) & = & -\left[a(p^{2}+m^{2})+b((p+k)^{2}+m^{2})+\tilde{a}(M^{2}+p^{2})+\tilde{b}(M^{2}+(p+k)^{2})\right]\\
 & = & -\left[(a+b+\tilde{a}+\tilde{b})p^{2}+(b+\tilde{b})(k^{2}+2p\cdot k)+(\tilde{a}+\tilde{b})M^{2}+(a+b)m^{2}\right]\\
 & = & -z\left[(p+(1-x)k)^{2}+\left[(1-x)-(1-x)^{2}\right]k^{2}+y\, M^{2}+(1-y)m^{2}\right]
\end{eqnarray*}
where in the last equality, the following change of variables has
been used,
\begin{eqnarray*}
z & = & (a+b+\tilde{a}+\tilde{b})\;,\; x=\frac{a+\tilde{a}}{z}\\
y & = & \frac{\tilde{a}+\tilde{b}}{z}\;,\; w=\frac{\tilde{b}}{z}
\end{eqnarray*}
the Jacobian and limits of integration in these new variables implying
that,
\[
\intop_{0}^{\infty}da\, db\, d\tilde{a}\, d\tilde{b}=\intop_{0}^{\infty}dz\intop_{0}^{1}dx\intop_{0}^{x}dy\intop_{0}^{y}(-z^{3})
\]
in terms of the variable $p'=\sqrt{z}(p+(1-x)k)$ the quantity $I_{\alpha}^{L}(k,m)$
is written as follows,
\begin{eqnarray*}
I_{\alpha}^{L}(k,m) & =- & \int d^{n}p'\; e^{-p'^{2}}\intop_{0}^{\infty}dz\; z^{3-\frac{n}{2}-2-2\alpha}\times\\
 &  & \intop_{0}^{1}dx\intop_{0}^{x}dy\intop_{0}^{y}dw\left[w(y-w)\right]^{-(1+\alpha)}e^{-z\left[x(1-x)k^{2}+y\, M^{2}+(1-y)m^{2}\right]}
\end{eqnarray*}
defining%
\footnote{As in the case of the tadpole, this definition of $d$ can be obtained
by matching the ultraviolet behavior of the integrands of $I_{\alpha}^{L}(k,m)$
and $I_{d}^{L}(k,m)$(the corresponding dimensionally regularized
integral)%
},
\[
\frac{d}{2}=\frac{n}{2}-2\alpha
\]
 and using (\ref{eq:power}) implies that,
\begin{eqnarray*}
I_{\alpha}^{L}(k,m) & = & -\frac{1}{\Gamma(-\alpha)^{2}}\int d^{n}p'\; e^{-p'^{2}}\Gamma(2-\frac{d}{2})\times\\
 &  & \intop_{0}^{1}dx\intop_{0}^{x}dy\intop_{0}^{y}dw\left[w(y-w)\right]^{-(1+\alpha)}\left[x(1-x)k^{2}+y\, M^{2}+(1-y)m^{2}\right]^{\frac{d}{2}-2}
\end{eqnarray*}
noting that,
\[
\int d^{n}p'\; e^{-p'^{2}}=\pi^{\frac{n}{2}}
\]
\[
\intop_{0}^{y}dw\left[w(y-w)\right]^{-(1+\alpha)}=\frac{2^{1+2\alpha}}{\sqrt{\pi}}\cos(\alpha\pi)\Gamma(-\alpha)\Gamma(\frac{1}{2}+\alpha)\, y^{-1-2\alpha}
\]
and,
\begin{eqnarray*}
\intop_{0}^{x}dy\, y^{-1-2\alpha}\,\left[x(1-x)k^{2}+y\, M^{2}+(1-y)m^{2}\right]^{\frac{d}{2}-2} & = & -\frac{x^{-2\alpha}}{2\alpha}(m^{2}+k^{2}x(1-x))^{\frac{d}{2}-2}\times\\
\,_{2}F_{1}(-2\alpha,2-4\alpha-\frac{d}{2},1-2\alpha,\frac{(m^{2}-M^{2})x}{m^{2}+k^{2}(1-x)x})
\end{eqnarray*}
where $\,_{2}F_{1}(a,b,c;z)$ denotes the hypergeometric function,
leads to the following expression for $I_{\alpha}^{L}(k,m)$,
\begin{eqnarray}
I_{\alpha}^{L}(k,m) & = & \pi^{\frac{n}{2}}\Gamma(2-\frac{d}{2})\frac{1}{\Gamma(-\alpha)}\intop_{0}^{1}dx\,\frac{2^{1+2\alpha}}{\sqrt{\pi}}\cos(\alpha\pi)\Gamma(\frac{1}{2}+\alpha)\times\nonumber \\
 &  & \frac{x^{-2\alpha}}{2\alpha}(m^{2}+k^{2}x(1-x))^{\frac{d}{2}-2}\,_{2}F_{1}(-2\alpha,2-\frac{d}{2},1-2\alpha,\frac{(m^{2}-M^{2})x}{m^{2}+k^{2}(1-x)x})\label{eq:ialfa-1}
\end{eqnarray}
For illustrative purposes, let us replace $m^{2}=M^{2}\frac{\alpha n}{1-2\alpha}$
by,
\[
\tilde{m}^{2}=m^{2}+m_{0}^{2}
\]
where $m_{0}^{2}$ is a constant additional mass. Taking the limit
$\alpha\to0$ of $I_{\alpha}^{L}(k,\tilde{m})$ gives,

\noindent 
\[
\lim_{\alpha\to0}I_{\alpha}^{L}(k,\tilde{m})=-\pi^{\frac{n}{2}}\Gamma(2-\frac{n}{2})\intop_{0}^{1}dx\,(m_{0}^{2}+k^{2}x(1-x))^{\frac{n}{2}}
\]
which, upon replacing $n$ by a complex number, is the dimensionally
regularized result of this diagram for a scalar field of mass $m_{0}$. 

A closed analytical result for the integral over $x$ in (\ref{eq:ialfa-1})
was not found. However, it can be shown numerically that $I_{\alpha}(k)$
is finite for all $\alpha$ such that $|\alpha|<1$, as in the case
of the tadpole diagram.

\section{Conclusions and outlook}

Conclusions and further research motivated by this work are summarized
in the series of remarks given below,
\begin{itemize}
\item It is useful to compare the approach presented in this work to the
usual dimensional regularization technique. In dimensional regularization,
each divergent integral is dealt with separately. There is no known
way of writing a Lagrangian leading to a perturbation theory where
all contributions turn out to be dimensionally regularized. As a consequence
it is not simple to make general statements about the dimensionally
regularized perturbation theory. Although a general proof has not
been given here, as regards ultraviolet power counting and comparison
with dimensional regularization , the perturbation theory obtained
taking as free Lagrangian the expression (\ref{eq:action}), leads
to finite contributions if $\alpha\neq0$. It would be misleading
to say that the present work gives a Lagrangian formulation of dimensional
regularization. A more precise statement is that it provides a regularization
scheme that is implementable at the Lagrangian level, and presents
the same singularity structure as dimensional regularization.
\item There is another related question which is considered of interest.
Are these theories only regularized theories? Do they make sense as
physical theories? At the level of perturbation theory this question
can be rephrased as follows: do they lead to quantum theories described
by a unitary $S$-matrix?
\item A very important feature of this approach is the fact that it is based
on a well-defined differential geometry. This allows to consider the
generalization of any field theory defined in differential geometric
terms to these deformed spaces. This includes gauge theories and gravity
theories. Of course the resulting theories deserve to be studied in
detail.
\item It is remarked that the approach presented in this work differs significantly
from the so-called non-commutative field theory\cite{dougnekra}(NCFT).
No non-commutativity of the coordinates is assumed. On the contrary,
non-commutativity enters at the level of the differential algebra
through the deformed choice of the Dirac operator. This difference
implies that this non-commutativity does not spoil the covariance
of adequately chosen field theories on these spaces. Furthermore,
as shown by the one loop calculations of the last section, the corresponding
contributions are finite, which is not in general the case for the
NCFTs. 
\end{itemize}
All the above remarks indicate that, from the point of view of physics,
further investigation of these theories is worth pursuing.

\section*{Appendix A: Junk Forms}

In this appendix it is shown that there are no junk forms for a non-zero
deformation parameter. To show this it is noted that a generic 1-form
can be written as,
\[
\omega^{(1)}=\sum_{I,J}\alpha_{JI}f_{J}df_{I}
\]
where the summation is over a complete basis $B=\{f_{I}\}$ for $\mathcal{A}$
and the $\alpha_{IJ}$ are numerical coefficients. Replacing (\ref{eq:dif})
in the last equation leads to,
\begin{equation}
\omega^{(1)}=\frac{1}{\Gamma(\alpha)}\int_{0}^{\infty}d\tau\,\tau^{\alpha-1}\sum_{I,J}\alpha_{JI}f_{J}\,\left[U(\tau)(Df_{I})+\left(U(\tau)f_{I}-f_{I}\, U(\tau)\right)D\right]\label{eq:1form}
\end{equation}
Junk 2-forms $\omega^{(2)}$are such that they can be written as the
differential of a vanishing 1-form, i.e.,
\[
\omega^{(2)}=d\omega^{(1)}\;,\;\;\omega^{(1)}=0
\]
thus the general expression for a vanishing 1-form is looked for.
From eq.(\ref{eq:1form}) this leads to the operatorial equation,
\[
0=\frac{1}{\Gamma(\alpha)}\int_{0}^{\infty}d\tau\,\tau^{\alpha-1}\sum_{I,J}\alpha_{JI}f_{J}\,\left[U(\tau)(Df_{I})+\left(U(\tau)f_{I}-f_{I}\, U(\tau)\right)D\right]
\]
this equation when applied to a constant spinor leads to,
\begin{equation}
0=\sum_{I,J}\alpha_{JI}f_{J}\, Df_{I}\label{eq:van}
\end{equation}
which is the same relation that appears for the $\alpha=0$ case.
The general solution is given by,
\begin{equation}
\omega^{(1)}=\sum_{I,J}\beta_{JI}f_{J}\,(2f_{I}df_{I}-d(f_{I}^{2}))\label{eq:coef}
\end{equation}
for arbitrary numerical coefficients $\beta_{JI}$. Using (\ref{eq:diftau})
gives,
\begin{eqnarray*}
d(f_{I}^{2}) & = & \frac{1}{\Gamma(\alpha)}\int_{0}^{\infty}d\tau\,\tau^{\alpha-1}d(f_{I}^{2})(\tau)\;\;,\; d(f_{I}^{2})(\tau)=U(\tau)2f_{I}(Df_{I})-[U(\tau),f_{I}^{2}]D\\
2f_{I}df_{I} & = & \frac{1}{\Gamma(\alpha)}\int_{0}^{\infty}d\tau\,\tau^{\alpha-1}d(f_{I}^{2})(\tau)\;\;,\;2f_{I}df_{I}(\tau)=2f_{I}\left\{ U(\tau)(Df_{I})-[U(\tau),f_{I}]D\right\} 
\end{eqnarray*}
thus,
\begin{eqnarray*}
2f_{I}df_{I}-d(f_{I}^{2})(\tau) & = & [U(\tau),f_{I}^{2}]D-2f_{I}[U(\tau),f_{I}]D-2[U(\tau),f_{I}](Df_{I})
\end{eqnarray*}
Applying equation (\ref{eq:coef}) to a constant spinor $\psi_{0}$
shows that in that case only the last term in the previous equation
contributes, therefore the equation $\omega^{(1)}=0$ leads to,
\[
0=\frac{1}{\Gamma(\alpha)}\int_{0}^{\infty}d\tau\,\tau^{\alpha-1}\sum_{I,J}\beta_{JI}f_{J}[U(\tau),f_{I}](Df_{I})\psi_{0}
\]
the linear independence of the basis $B=\{f_{J}\}$ implying that,
\begin{equation}
0=\frac{1}{\Gamma(\alpha)}\int_{0}^{\infty}d\tau\,\tau^{\alpha-1}\sum_{I}\beta_{JI}[U(\tau),f_{I}](Df_{I})\psi_{0}\;\;,\forall J\label{eq:01form}
\end{equation}
Next the expansion of the quantity $[U(\tau),f_{I}](Df_{I})$ in the
basis $B$ is considered, i.e.,
\[
[U(\tau),f_{I}](Df_{I})=\sum_{K}\alpha_{K}^{I}(\tau)\, f_{K}
\]
the coefficients $\alpha_{K}^{I}$ being given by,
\[
\alpha_{K}^{I}(\tau)=\int_{x}f_{K}^{*}[U(\tau),f_{I}](Df_{I})
\]
at this stage it is convenient to use the Fourier basis $f_{I}=e^{iI\cdot x}\;\;,I\in\mathbb{Z}^{n}$,
which are eigenstates of $D^{2}.$ Noting that,
\[
e^{iI\cdot(x-2\tau\partial)}=e^{iI\cdot x}e^{-i2\tau I\cdot\partial}e^{\tau I\cdot I}
\]
leads to,
\begin{eqnarray}
\alpha_{K}^{I}(\tau) & = & \int_{x}e^{-iK\cdot x}(e^{iI\cdot(x-2\tau\partial)}-1)U(\tau)(De^{iI\cdot x})\nonumber \\
 & = & \int_{x}e^{-iK\cdot x}(e^{iI\cdot(x-2\tau\partial)}-1)e^{iI\cdot x}e^{-\tau(1+I^{2})}(-\gamma\cdot I)\nonumber \\
 & = & \delta(K-2I)C(\tau,I)\label{eq:allfak}
\end{eqnarray}
where the matrix $C(\tau,I)$ is given by,
\[
C(\tau,I)=(e^{-3\tau I^{2}}-1)e^{-\tau(1+I^{2})}(-\gamma\cdot I)
\]
 replacing in (\ref{eq:01form}) leads to,
\[
0=\frac{1}{\Gamma(\alpha)}\int_{0}^{\infty}d\tau\,\tau^{\alpha-1}\sum_{I,K}\beta_{JI}\alpha_{K}^{I}(\tau)\, f_{K}\psi_{0}\;\;,\forall j
\]
which taken into account the linear independence of the basis $B$
and replacing (\ref{eq:allfak}), implies that,
\begin{eqnarray*}
0 & = & \frac{1}{\Gamma(\alpha)}\int_{0}^{\infty}d\tau\,\tau^{\alpha-1}\beta_{JI}C(\tau,I)\;\;,\forall\, I,J\\
 & = & \beta_{JI}\left[(1+4I^{2})^{\alpha}-(1+I^{2})^{\alpha}\right](-\gamma\cdot I)
\end{eqnarray*}
which is solved by,
\[
\beta_{IJ}=0\;\;,\forall\, J,I\neq0
\]
the case $\beta_{0J}\neq0$ is trivial since anyhow in that case $\omega^{(1)}=0$.

\section*{Appendix B: Evaluation of the Diximiers trace}

As shown in section \ref{sec:The-scalar-field} the actions to be
evaluated are,
\begin{eqnarray*}
S_{k} & = & tr_{\omega}\left\{ [U_{\alpha}D,\phi][DU_{\alpha},\phi^{*}]|D|^{-n-2k}\right\} \;\;,\; U_{\alpha}=(1+D^{2})^{-\alpha}\\
 & = & tr_{\omega}\left\{ \left(2\phi DU_{\alpha}\phi^{*}U_{\alpha}D-\phi\phi^{*}(U_{\alpha}D)^{2}-\phi(U_{\alpha}D)^{2}\phi^{*}\right)|D|^{-n-2k}\right\} =tr_{\omega}\left\{ A_{k}\right\} 
\end{eqnarray*}
these Diximier traces will be evaluated using their expression as
Wodzicki residues,
\[
S_{k}=\frac{1}{n(2\pi)^{n}}\int_{S^{*}T^{n}}tr\sigma_{-n}^{A_{k}}(x,\xi)
\]
where $\sigma_{-n}^{A_{k}}(x,\xi)$ denotes the term of order $-n$
of the symbol of the operator $A_{k}$ , $(x,\xi)$ denote coordinates
over the unit co-sphere on the cotangent bundle of $T^{n}$, so that
$\int_{S^{*}T^{n}}=\int_{x}\int_{\xi}d\Omega_{n-1}$where $d\Omega_{n-1}$
is the volume element in the sphere $S_{n-1}$. The trace is taken
over the spin space and the symbol is defined by,
\[
\sigma^{A_{k}}(x,\xi)=e^{-ix\cdot\xi}A_{k}\; e^{ix\cdot\xi}
\]
so that,
\begin{align*}
\sigma^{A_{k}}(x,\xi) & =\left(\right.2e^{-ix\cdot\xi}\phi DU_{\alpha}\phi^{*}e^{ix\cdot\xi}(-\gamma\cdot\xi)(1+|\xi|^{2})^{-\alpha}\\
 & -\phi\phi^{*}(1+|\xi|^{2})^{-\alpha}|\xi|^{2}-\phi(DU_{\alpha})^{2}\phi^{*}\left.\right)|\xi|^{-n-2k}
\end{align*}
using that,
\[
e^{-ix\cdot\xi}DU_{\alpha}e^{ix\cdot\xi}=(D-\gamma\cdot\xi)(1+(D-\gamma\cdot\xi)^{2})^{-\alpha}
\]
ignoring terms that vanish when integrating over $|\xi|=1$ and evaluating
the trace, leads to, 
\begin{eqnarray*}
tr\sigma^{A_{k}}(x,\xi) & = & 2^{[\frac{n}{2}]}\left(2\phi(1+(D-\xi)^{2})^{-\alpha}\phi^{*}(1+|\xi|^{2})^{-\alpha}|\xi|^{-n-2k+2}\right.\\
 &  & -\phi\phi^{*}(1+|\xi|^{2})^{-2\alpha}|\xi|^{-n-2k+2}\\
 &  & \left.-\phi[D^{2}+\xi^{2}](1+(D-\xi)^{2})^{-2\alpha}\phi^{*}|\xi|^{-n-2k}\right)
\end{eqnarray*}
for $\alpha\neq0$ the first two terms do not contribute to the term
of order $-n$ of the symbol because they include the factor,
\[
(1+|\xi|^{2})^{-\alpha}=\sum_{k=0}^{\infty}\left(\begin{array}{c}
-\alpha\\
k
\end{array}\right)|\xi|^{-2(\alpha+k)}
\]
which decrease the order of the corresponding terms by at least $-2\alpha$.
The last term gives two non-vanishing contributions. One coming from
the term $D^{2}$ inside the square bracket, which contributes to
the term of order $-n$ of the symbol only when $k=0$. And the other
coming form the $\xi^{2}$ inside the square bracket, which contributes
to the term of order $-n$ of the symbol only when $k=1$. Thus,
\begin{eqnarray*}
\sigma_{-n}^{A_{0}}(x,1) & = & -\phi D^{2}(1+D{}^{2})^{-2\alpha}\phi^{*}\\
\sigma_{-n}^{A_{1}}(x,1) & = & -\phi(1+D{}^{2})^{-2\alpha}\phi^{*}
\end{eqnarray*}
hence,
\begin{eqnarray*}
S_{0} & = & -\frac{2^{[\frac{n}{2}]}V_{S^{n-1}}}{n(2\pi)^{n}}\int_{T^{n}}\phi D^{2}(1+D{}^{2})^{-2\alpha}\phi^{*}\\
S_{1} & = & -\frac{2^{[\frac{n}{2}]}V_{S^{n-1}}}{n(2\pi)^{n}}\frac{\alpha n}{1-2\alpha}\int_{T^{n}}\phi(1+D{}^{2})^{-2\alpha}\phi^{*}
\end{eqnarray*}
leading to,
\[
S=-\frac{2^{[\frac{n}{2}]}V_{S^{n-1}}}{n(2\pi)^{n}}\int_{T^{n}}\phi(D^{2}+\frac{\alpha n}{1-2\alpha})(1+D{}^{2})^{-2\alpha}\phi^{*}
\]
where $V_{S^{n-1}}$ denotes the surface of the sphere $S^{n-1}$
given by, 
\[
V_{S^{n-1}}=\int d\Omega_{n-1}=\frac{2\pi^{\frac{n}{2}}}{\Gamma(\frac{n}{2})}
\]

\section*{Acknowledgments}

I want to express my gratitude to S. Capriotti, C. Fosco and S. Grillo
for valuable comments and suggestions.

\end{document}